\documentclass[letterpaper, 10pt, conference]{ieeeconf}
\usepackage{amsmath,amssymb,amsfonts}
\usepackage{algorithmic}
\usepackage{algorithm}
\usepackage{graphicx}
\usepackage{textcomp}
\usepackage{booktabs}
\usepackage{multirow}
\usepackage{titlesec}
\titlespacing*{\subsubsection}{0pt}{0.5\baselineskip}{0.2\baselineskip}
\usepackage{nameref}
\titlespacing*{\subsection}{0pt}{0.5\baselineskip}{0.25\baselineskip}
\makeatletter
\let\NAT@parse\undefined
\makeatother
\usepackage[hyperfootnotes=false]{hyperref}
\usepackage{float}

\usepackage{xcolor}
\usepackage{placeins}
\usepackage{subcaption}
\usepackage{caption}
\usepackage{stfloats}

\newtheorem{theorem}{Theorem}
\newtheorem{lemma}{Lemma}
\newtheorem{definition}{Definition}
\newtheorem{assumption}{Assumption}

\IEEEoverridecommandlockouts
\title{\LARGE \bf Safe Deep Reinforcement Learning for\\Energy-Efficient HVAC Control in Multi-Zone Residential Buildings}

\author{Oussama Ziadi$^{1}$, Abdelilah Rochd$^{2}$, Samir Idrissi Kaitouni$^{2}$,\\
Mohamed Oualid Mghazli$^{2}$, Adnane Saoud$^{1}$%
\thanks{$^{1}$College of Computing, Mohammed VI Polytechnic University (UM6P), Ben Guerir, Morocco. {\tt\small \{oussama.ziadi, adnane.saoud\}@um6p.ma}}%
\thanks{$^{2}$Green Energy Park, Route Regional Kelaa, Km3, R206, Morocco. {\tt\small \{rochd, kaitouni, mghazli\}@greenenergypark.ma}}%
}

\begin{document}

\maketitle

\begin{abstract}
HVAC systems represent a major share of building energy consumption.
Traditional control strategies are limited in coordinating
energy-comfort tradeoffs across multiple zones simultaneously.
Reinforcement learning (RL) offers adaptive, data-driven control that
optimizes performance over time. However, deploying learned neural
network controllers in safety-critical building systems remains
challenging due to lack of formal safety guarantees. We propose a
safety-certified deep RL framework for multi-zone residential
HVAC control. Proximal Policy Optimization (PPO) and Soft Actor-Critic
(SAC) agents are trained in an EnergyPlus/Sinergym simulation to
minimize energy consumption while maintaining thermal comfort.
Post-training safety certification is performed on the PPO
policy using Lipschitz-based forward invariance analysis,
building on existing tools for the computation of Lipschitz constants
for neural networks, to guarantee constraint satisfaction.
Both agents are evaluated over an annual simulation cycle in an
eight-zone variable refrigerant flow (VRF) testbed. The PPO
agent achieves 67\% comfort violation reduction compared to
rule-based control, while the SAC agent achieves 27.6\% energy
savings. The PPO policy satisfies formal safety certification
with a margin of $2.003^\circ$C. These results demonstrate the
feasibility of combining reinforcement learning with
post-training safety verification for multi-zone building
control.
\end{abstract}

\section{Introduction}

The building sector accounts for more than 40\% of global energy use,
with Heating, Ventilation, and Air Conditioning (HVAC) systems
representing the dominant share~\cite{perez2008review}. The dual
objectives of minimizing energy use and maintaining occupant thermal
comfort define a multi-objective optimization problem complicated by
nonlinear thermodynamics, stochastic occupancy, and weather
variability~\cite{wang2020reinforcement}. Traditional control
techniques such as proportional-integral-derivative (PID) 
typically regulate individual zone setpoints without
supervisory-level coordination of competing objectives across
zones~\cite{oldewurtel2012use}. Model predictive control (MPC)
can optimize across zones and has a long track record in industrial
HVAC applications~\cite{oldewurtel2012use}, but requires
system-specific thermal models that must be developed and calibrated
for each building configuration. Reinforcement learning (RL) offers a
model-free alternative that learns control policies directly from
building interaction~\cite{vazquez2019reinforcement}.

Recent work has demonstrated RL effectiveness for HVAC control.
\cite{wei2017deep} applied deep Q-networks to single-zone temperature
control; \cite{zhang2019whole} extended RL to multi-zone radiant
heating. Variable refrigerant flow (VRF) systems constitute a specific
HVAC technology enabling simultaneous heating/cooling across multiple
zones through modulated refrigerant
distribution~\cite{aynur2010variable}. Prior work on RL-based VRF
control~\cite{yoon2019performance} considered single-building
settings; extension to multi-zone residential testbeds with
heterogeneous thermal characteristics remains underexplored.

A critical limitation of neural network controllers is the absence of
formal safety guarantees. While RL agents may achieve favorable
average performance, worst-case constraint violations pose risks in
occupied buildings~\cite{berkenkamp2017safe,garcia2015comprehensive},
manifesting as thermal discomfort, equipment damage, or energy
waste during extreme weather events.
Post-training verification using Lipschitz continuity
bounds~\cite{fazlyab2019efficient} offers a computationally tractable
approach: the Lipschitz constants of the policy and the dynamics
bound how far the closed-loop state can drift per step, and safety
is established by verifying forward invariance of the safe
set~\cite{jouret2023safety,kazemi2024datadriven}. We apply this
framework to multi-zone HVAC control for the first time in the
literature: we estimate the Lipschitz constants of the building
thermal dynamics from trajectory data, compute the Lipschitz constant
of the policy network using LipSDP~\cite{fazlyab2019efficient}, and
certify safety of the closed-loop system via invariance analysis on
the temperature subspace.

This paper makes the following contributions:
\begin{itemize}
    \item An 8-zone VRF control environment based on the Green Energy
    Park platform, implemented in
    EnergyPlus~\cite{energyplus_docs} and interfaced via
    Sinergym~\cite{jimenez2024sinergym};
    \item Training of PPO and SAC agents for multi-zone setpoint
    control with comparison against a rule-based baseline;
    \item Development of a Lipschitz-based post-training safety
    certification~\cite{fazlyab2019efficient} framework for HVAC control;
    \item Simulation-based validation establishing certified
    forward invariance with safety margin of
    $2.003^\circ$C under bounded disturbances.
\end{itemize}

\section{Methodology}
\label{sec:methodology}

Training and evaluation of the RL agents are conducted through
co-simulation between Sinergym~\cite{jimenez2024sinergym} and
EnergyPlus~\cite{energyplus_docs}. 

\subsection{Building Simulation Environment}

The simulation model is based on a residential multi-building
testbed located at the Green Energy Park research platform in
Ben~Guerir, Morocco. The eight houses were constructed for the
Solar Decathlon Africa 2019 and differ in architecture, orientation,
and envelope materials~\cite{gep_reference}. The thermal model is
implemented in EnergyPlus~23.2 using the system curve-based VRF
model, interfaced through Sinergym~\cite{jimenez2024sinergym} and
coupled with Stable-Baselines3~\cite{sb3} for RL training.

The testbed comprises 8 controlled thermal zones with a total
conditioned floor area of 256.0~m$^2$. Each zone is served by a VRF
terminal unit connected to a centralized outdoor unit. The
EnergyPlus VRF model captures the essential thermodynamic interactions
(zone heat balance, refrigerant flow modulation, shared outdoor unit
capacity allocation) but uses idealized performance curves that do not
reflect manufacturer-specific constraints on simultaneous heating and
cooling or detailed refrigerant-side dynamics~\cite{aynur2010variable};
accordingly, the specific numerical results reflect the fidelity of the simulation model.

\subsubsection{Internal Heat Gains}
Table~\ref{tab:zone_loads} summarizes zone-specific parameters.
Equipment power density ranges from 6 to 15~W/m$^2$; lighting power
density from 6 to 20~W/m$^2$. The current model captures sensible heat gains; latent loads are outside the scope of this study.

\begin{table}[!t]
\centering
\caption{Internal Heat Gains for Controlled Zones}
\label{tab:zone_loads}
\begin{tabular}{lcccc}
\toprule
\textbf{Zone} & \textbf{Area} & \textbf{Occ.} & \textbf{Light} & \textbf{Equip} \\
 & (m$^2$) & & (W) & (W) \\
\midrule
ZONE 1 & 35.4 & 6 & 283 & 496 \\
ZONE 2 & 48.9 & 4 & 489 & 684 \\
ZONE 3 & 52.0 & 3 & 623 & 779 \\
ZONE 4 & 11.7 & 4 & 70 & 105 \\
ZONE 5 & 46.2 & 3 & 370 & 277 \\
ZONE 6 & 16.1 & 8 & 323 & 242 \\
ZONE 7 & 24.0 & 4 & 240 & 336 \\
ZONE 8 & 21.7 & 3 & 261 & 152 \\
\midrule
\textbf{Total} & \textbf{256.0} & \textbf{35} & \textbf{2659} & \textbf{3071} \\
\bottomrule
\end{tabular}
\end{table}

The testbed follows residential occupancy patterns in
Table~\ref{tab:schedules}. The schedules are deterministic time-of-day profiles, following standard practice in residential building simulation~\cite{manjavacas2024experimental}.

\begin{table}[!t]
\centering
\caption{Occupant, Equipment, and Lighting Schedules (Fraction of Full Load)}
\label{tab:schedules}
\begin{tabular}{l|cccc}
\toprule
 & \textbf{6--8h} & \textbf{8--17h} & \textbf{17--22h} & \textbf{22--6h} \\
\midrule
Occupancy & 80\% & 20\% & 95\% & 100\% \\
Equipment & 35\% & 25\% & 80\% & 30\% \\
Lighting & 20\% & 10\% & 90\% & 30\% \\
\bottomrule
\end{tabular}
\end{table}

\subsubsection{Weather Data}
Training and evaluation use Typical Meteorological Year data from
Ben~Guerir, Morocco. The simulation timestep is 10 minutes, yielding
52,560 decision points per annual episode.

\subsection{Control Problem Formulation}

Let $\Sigma$ denote the building thermal dynamics:
\begin{equation}
\Sigma: \quad x_{t+1} = f(x_t, u_t, w_t)
\label{eq:dynamics_model}
\end{equation}
where $x_t \in \mathcal{X} \subseteq \mathbb{R}^{n}$ is the state,
$u_t \in \mathcal{U} \subseteq \mathbb{R}^{m}$ is the control input
(temperature setpoints), $w_t \in \mathcal{W}$ represents exogenous
disturbances, and $f$ is the nonlinear transition function implemented
by EnergyPlus.

\subsubsection{State Space}
The state vector $x_t \in \mathbb{R}^{84}$ consists of temporal
features (month, day, hour), weather variables (outdoor temperature,
humidity, wind speed/direction, diffuse and direct solar radiation),
aggregate HVAC power (total, cooling, heating), and per-zone
measurements for each of the 8~zones (air temperature, humidity,
heating/cooling setpoints, terminal unit power, occupant count,
lighting and equipment loads).

\subsubsection{Action Space}
The action vector $a_t \in \mathbb{R}^{16}$ specifies heating and
cooling setpoints for each zone:
\begin{equation}
a_t = [T_{1,\text{htg}}^\text{sp}, T_{1,\text{clg}}^\text{sp}, \ldots, T_{8,\text{htg}}^\text{sp}, T_{8,\text{clg}}^\text{sp}]^\top
\end{equation}
with $T_{i,\text{htg}}^\text{sp} \in [15, 23]^\circ$C and
$T_{i,\text{clg}}^\text{sp} \in [23, 30]^\circ$C. These ranges follow standard VRF supervisory practice: the
heating ceiling and cooling floor meet at $23^\circ$C, forming a
deadband that prevents simultaneous heating and cooling. The comfort
band $[20, 26]^\circ$C lies within both ranges, ensuring the agent
can always actuate in either direction before a comfort violation
occurs~\cite{aynur2010variable}.

\subsubsection{Reward Function}
The reward balances energy consumption and thermal comfort:
\begin{equation}
r_t = -\omega \lambda_E P_{\text{HVAC},t} - (1-\omega) \lambda_T \sum_{i=1}^{8} c(T_{i,t})
\label{eq:reward}
\end{equation}
where $\omega \in [0,1]$ balances energy and comfort;
$\lambda_E, \lambda_T$ are scaling constants; and the comfort penalty
is:
\begin{equation}
c(T) = \begin{cases}
T_\text{min} - T & \text{if } T < T_\text{min} \\
0 & \text{if } T_\text{min} \leq T \leq T_\text{max} \\
T - T_\text{max} & \text{if } T > T_\text{max}
\end{cases}
\label{eq:comfort}
\end{equation}
with $T_\text{min} = 20^\circ$C, $T_\text{max} = 26^\circ$C.
The linear formulation penalizes deviations in proportion to
their magnitude, providing a continuous gradient signal suited to
policy optimization. While richer comfort models such as the Predicted
Mean Vote (PMV) index capture humidity and air velocity effects, they
require additional sensor inputs not available in the current
testbed; the linear penalty is a standard choice in RL-based building
control~\cite{manjavacas2024experimental,biemann2021experimental}.

\subsection{Controllers}

\subsubsection{Baseline Controller}
The RL controllers are compared to a rule-based controller (RBC),
following~\cite{manjavacas2024experimental}. The RBC implements
occupancy-responsive setpoint adjustment with a deadband
$\Delta = 0.5^\circ$C: during occupied periods, heating setpoints
increase when $T_i < T_{\min} - \Delta$ and cooling setpoints decrease
when $T_i > T_{\max} + \Delta$; during unoccupied periods,
energy-saving setback temperatures are applied
($T_\text{htg}^\text{sp} = 18^\circ$C,
$T_\text{clg}^\text{sp} = 27^\circ$C). This reactive baseline operates on the same observation variables, action
bounds, and comfort band as the RL agents, ensuring a fair
comparison: all controllers see the same inputs and are subject to
the same constraints. This makes the RBC a controlled reference
point for evaluating what the learning algorithms bring beyond
simple reactive logic~\cite{manjavacas2024experimental}.

\subsubsection{Reinforcement Learning Agents}
We employ PPO~\cite{schulman2017proximal} and
SAC~\cite{haarnoja2018soft} for continuous setpoint control. PPO is
on-policy with clipped updates for stability; SAC is off-policy with
entropy regularization for exploration. Both use feedforward networks
(two hidden layers, 128 units, ReLU) implemented in
Stable-Baselines3~\cite{sb3}. Agents were trained for 50 episodes with
stochastic weather noise. Hyperparameters
(Table~\ref{tab:hyperparams}) were tuned via
Optuna~\cite{optuna}.

\begin{table}[!t]
\centering
\caption{Hyperparameters (Optuna-tuned)}
\label{tab:hyperparams}
\begin{tabular}{lcc}
\toprule
\textbf{Parameter} & \textbf{PPO} & \textbf{SAC} \\
\midrule
Learning rate & $3 \times 10^{-4}$ & $3 \times 10^{-4}$ \\
Batch size & 64 & 256 \\
Discount $\gamma$ & 0.99 & 0.99 \\
GAE $\lambda$ & 0.95 & -- \\
Entropy coef. & 0.01 & auto \\
Target smoothing $\tau$ & -- & 0.005 \\
Buffer size & -- & $10^6$ \\
Training steps & $5 \times 10^5$ & $5 \times 10^5$ \\
\bottomrule
\end{tabular}
\end{table}

\section{Safety Certification}
\label{sec:safety_certification}

This section presents a post-training verification framework
that certifies whether a trained policy maintains zone temperatures
within comfort bounds. Existing Lipschitz-based verification methods
either address the neural network in
isolation~\cite{fazlyab2019efficient} or assume analytically known
dynamics~\cite{jouret2023safety}. In our setting, EnergyPlus solves
coupled nonlinear heat balance equations numerically, so that
Lipschitz constants of the thermal dynamics cannot be derived
analytically and must instead be estimated from trajectory data. The
framework therefore combines LipSDP~\cite{fazlyab2019efficient} for
the computation of the Lipschitz constant of the policy network
with data-driven Lipschitz estimation for the
dynamics~\cite{kazemi2024datadriven}, and performs projected
safety certification: invariance is verified on the temperature
subspace where comfort constraints are defined, rather than on the
full state-space.

\subsection{Problem Statement}

When the control inputs are generated by the trained policy
$\Pi_\theta: \mathcal{X} \to \mathcal{U}$, the closed-loop dynamics
become:
\begin{equation}
x_{t+1} = f(x_t, \Pi_\theta(x_t), w_t) \triangleq F_\theta(x_t, w_t)
\label{eq:closed_loop}
\end{equation}

\begin{definition}[Safe Set]
$\mathcal{S} = \{x \in \mathcal{X} : T_{\min} \leq z_i \leq T_{\max},\; \forall\, i = 1, \ldots, 8\}$.
\end{definition}

\begin{definition}[Robust Controlled Invariance~\cite{blanchini1999set}]
A set $K \subseteq \mathcal{X}$ is robust controlled invariant for
$(\Sigma, X, U, W)$ if $K \subseteq X$ and there exists
$\kappa: X \to U$ such that
$\Phi_\kappa(t, x_0, \mathbf{w}) \in K$ for all $x_0 \in K$,
all $\mathbf{w}: \mathbb{N}_{\geq 0} \to W$, and all $t \geq 0$.
An equivalent one-step characterization~\cite{saoud2021composition}:
$K$ is robust controlled invariant iff $K \subseteq X$ and
$\forall\, x \in K,\; \exists\, u \in U$ s.t.\
$f(x, u, W) \subseteq K$.
\end{definition}

Our goal is to certify that $\Pi_\theta$ renders $\mathcal{S}$
robustly forward invariant.

\subsection{Lipschitz-Based Certification}

The state decomposes as $x = (z, \bar{z})$ where
$z \in \mathbb{R}^8$ contains zone temperatures and
$\bar{z} \in \mathbb{R}^{76}$ the remaining variables.
We define the projection operator
$\mathcal{P}_z: \mathcal{X} \to \mathbb{R}^8$ that extracts the
temperature components:
\begin{equation}
\mathcal{P}_z(x) = \mathcal{P}_z(z, \bar{z}) = z.
\label{eq:projection}
\end{equation}
This projection allows us to focus the safety analysis on the
temperature subspace where comfort constraints are defined.

\begin{assumption}[Lipschitz Continuity]
\label{assump:lip}
There exist nonnegative constants $L_\theta, L_z, L_{\bar{z}}, L_u, L_w$ such that the following hold.

\textbf{Policy:} the neural network policy
$\Pi_\theta: \mathcal{X} \to \mathcal{U}$ satisfies
\begin{equation*}
\|\Pi_\theta(x) - \Pi_\theta(x')\| \leq L_\theta \|x - x'\|,
\quad \forall\, x, x' \in \mathcal{X}.
\end{equation*}
\textbf{Dynamics:} the transition map $f$ satisfies the following
channel-wise Lipschitz conditions:
\begin{itemize}
    \item \textbf{Temperature channel:}
    \begin{equation*}
    \|\mathcal{P}_z\bigl(f(z,\bar{z},u,w) - f(z',\bar{z},u,w)\bigr)\|
    \leq L_z \|z - z'\|,
    \end{equation*}
    for all $z, z' \in \mathbb{R}^8$, $\bar{z} \in \mathbb{R}^{76}$,
    $u \in \mathcal{U}$, $w \in \mathcal{W}$;
    \item \textbf{Auxiliary channel:}
    \begin{equation*}
    \|\mathcal{P}_z\bigl(f(z,\bar{z},u,w) - f(z,\bar{z}',u,w)\bigr)\|
    \leq L_{\bar{z}} \|\bar{z} - \bar{z}'\|,
    \end{equation*}
    for all $z \in \mathbb{R}^8$,
    $\bar{z}, \bar{z}' \in \mathbb{R}^{76}$, $u \in \mathcal{U}$,
    $w \in \mathcal{W}$;
    \item \textbf{Control channel:}
    \begin{equation*}
    \|\mathcal{P}_z\bigl(f(x,u,w) - f(x,u',w)\bigr)\|
    \leq L_u \|u - u'\|,
    \end{equation*}
    for all $x \in \mathcal{X}$, $u, u' \in \mathcal{U}$,
    $w \in \mathcal{W}$;
    \item \textbf{Disturbance channel:}
    \begin{equation*}
    \|\mathcal{P}_z\bigl(f(x,u,w) - f(x,u,w')\bigr)\|
    \leq L_w \|w - w'\|,
    \end{equation*}
    for all $x \in \mathcal{X}$, $u \in \mathcal{U}$,
    $w, w' \in \mathcal{W}$.
\end{itemize}
\end{assumption}

The decomposition separates sensitivity into distinct channels:
$L_z$ captures thermal inertia (how current temperatures affect
future temperatures); $L_{\bar{z}}$ captures the influence of weather
and occupancy on temperature evolution; $L_w$ captures the
sensitivity to disturbance variations; and $L_u L_\theta$ captures
how the control policy responds to state changes through the
composition of the policy's input sensitivity and the dynamics'
response to control inputs.

\begin{lemma}[Closed-Loop Sensitivity]
\label{lemma:sensitivity}
Under Assumption~\ref{assump:lip}, for all $x, x' \in \mathcal{X}$
and $w, w' \in \mathcal{W}$:
\begin{equation}
\begin{aligned}
\|\mathcal{P}_z(F_\theta(x, w) &- F_\theta(x', w'))\| \\
&\leq L_z\|z - z'\| + L_{\bar{z}}\|\bar{z} - \bar{z}'\| \\
&\quad + L_u L_\theta \|x - x'\| + L_w \|w - w'\|
\end{aligned}
\label{eq:closed_loop_bound}
\end{equation}
\end{lemma}

\textit{Proof:}
Let $x = (z, \bar{z})$, $x' = (z', \bar{z}')$ with control actions
$u = \Pi_\theta(x)$ and $u' = \Pi_\theta(x')$, and let
$w, w' \in \mathcal{W}$. We introduce three intermediate states that
vary one channel at a time:
\begin{align*}
y_1 &= f(z', \bar{z}, u, w), \\
y_2 &= f(z', \bar{z}', u, w), \\
y_3 &= f(z', \bar{z}', u', w).
\end{align*}
By the triangle inequality,
\begin{align*}
\|\mathcal{P}_z(F_\theta(x,w) - F_\theta(x',w'))\|
&\leq \|\mathcal{P}_z(f(z,\bar{z},u,w) - y_1)\| \\
&\quad + \|\mathcal{P}_z(y_1 - y_2)\| \\
&\quad + \|\mathcal{P}_z(y_2 - y_3)\| \\
&\quad + \|\mathcal{P}_z(y_3 - f(z',\bar{z}',u',w'))\|.
\end{align*}
Applying the four channel-wise bounds of Assumption~\ref{assump:lip}
to each term in order; temperature, auxiliary, control, and
disturbance yields
\begin{align*}
\|\mathcal{P}_z(F_\theta(x,w) - F_\theta(x',w'))\|
&\leq L_z \|z - z'\| + L_{\bar{z}} \|\bar{z} - \bar{z}'\| \\
&\quad + L_u \|u - u'\| + L_w \|w - w'\|.
\end{align*}
Finally, the policy Lipschitz property gives
$\|u - u'\| = \|\Pi_\theta(x) - \Pi_\theta(x')\| \leq L_\theta\|x - x'\|$,
which substituted into the third term completes the bound.
\hfill$\blacksquare$

\subsection{Safety Certification via Data Coverage}

Let $\mathcal{D} = \{(x_j, y_j)\}_{j=1}^N$ be a trajectory dataset
where $y_j = F_\theta(x_j, w_j)$. The dataset covers the operating
region if for any $x \in \mathcal{X}$ there exists $x_j \in \mathcal{D}$
with $\|x - x_j\| \leq \varepsilon_x$,
$\|z - z_j\| \leq \varepsilon_z$, and
$\|\bar{z} - \bar{z}_j\| \leq \varepsilon_{\bar{z}}$.

\begin{theorem}[Safety Certification]
\label{thm:safety}
Let Assumption~\ref{assump:lip} hold and let $\mathcal{D}$ satisfy
the covering property. Define the safety buffer:
\begin{equation}
r = L_z \varepsilon_z + L_{\bar{z}} \varepsilon_{\bar{z}} + L_u L_\theta \varepsilon_x + L_w \varepsilon_w
\label{eq:buffer}
\end{equation}
If (i)~$r < \delta$ where $\delta = (T_{\max} - T_{\min})/2$, and
(ii)~$\mathcal{P}_z(y_j) \in [T_{\min}+r,\; T_{\max}-r]^8$ for
all $j = 1,\ldots,N$, then $\mathcal{S}$ is robustly forward
invariant under $\Pi_\theta$.
\end{theorem}

\textit{Proof:} For any $x \in \mathcal{S}$ and
$w \in \mathcal{W}$, the covering property provides
$(x_j, y_j) \in \mathcal{D}$ within the covering radii.
Lemma~\ref{lemma:sensitivity} gives
$\|\mathcal{P}_z(F_\theta(x,w) - y_j)\| \leq r$. Since
$\mathcal{P}_z(y_j) \in [T_{\min}+r,\, T_{\max}-r]^8$ by~(ii), a
perturbation bounded by $r$ yields
$\mathcal{P}_z(F_\theta(x,w)) \in [T_{\min},T_{\max}]^8$, hence
$F_\theta(x,w) \in \mathcal{S}$. \hfill$\blacksquare$

\subsection{Estimation of Lipschitz Constants and Covering Radii}

\textbf{Policy Lipschitz constant.} We compute $L_\theta$ using
LipSDP~\cite{fazlyab2019efficient}, a semidefinite programming
approach that exploits the slope-restricted nature of ReLU activations
($\sigma'(x) \in [0,1]$) to encode quadratic constraints on the
network's input-output mapping. This yields tighter bounds than the
naive product of per-layer spectral norms
$\bar{L}_\theta = \prod_{l} \sigma_{\max}(W_l)$, which does not
account for inter-layer coupling.

\textbf{Dynamics Lipschitz constants.} Since the EnergyPlus
thermal dynamics involve coupled nonlinear subsystems from which
closed-form constants cannot be extracted, we estimate $L_u$, $L_z$,
$L_{\bar{z}}$, $L_w$ from trajectory data following the
data-driven approach
of~\cite{wood2017experiment,kazemi2024datadriven}. For each channel,
trajectory pairs differing primarily in the corresponding argument are
identified. For example, the control channel constant is estimated as:
\begin{equation}
\hat{L}_u = \max_{\substack{(x_t, x_{t'}) \in \mathcal{D},\;
    \|x_t - x_{t'}\| < \epsilon \\ u_t \neq u_{t'}}}
    \frac{\|\mathcal{P}_z(x_{t+1} - x_{t'+1})\|}{\|u_t - u_{t'}\|}
\end{equation}
where $\epsilon$ is a small tolerance for state similarity. Analogous
formulations apply for $\hat{L}_z$, $\hat{L}_{\bar{z}}$, and
$\hat{L}_w$, each isolating the effect of a single channel.

\textbf{Covering radii.} The radii $\varepsilon_x$, $\varepsilon_z$,
$\varepsilon_{\bar{z}}$, $\varepsilon_w$ quantify how densely the
dataset covers each subspace. They are computed as maximum
nearest-neighbor distances:
$\varepsilon_x = \max_{j}\min_{k \neq j} \|x_j - x_k\|$,
and analogously for $\varepsilon_z$, $\varepsilon_{\bar{z}}$,
$\varepsilon_w$. The maximum nearest-neighbor distance is the smallest radius that
guarantees every state lies within $\varepsilon$ of some data point,
making the certification valid across the entire observed operating
region.

Algorithm~\ref{alg:certification} summarizes the certification
procedure: compute Lipschitz constants, evaluate covering radii,
assemble the safety buffer, and verify that conditions~(i)--(ii) of
Theorem~\ref{thm:safety} hold.

\begin{algorithm}[!t]
\caption{Safety Certification via Lipschitz Analysis}
\label{alg:certification}
\begin{algorithmic}[1]
\footnotesize
\REQUIRE Trained policy $\Pi_\theta$, trajectory data $\mathcal{D}$, comfort bounds $[T_{\min}, T_{\max}]$
\ENSURE Safety verdict and margin
\STATE Compute $L_\theta$ via LipSDP on policy network weights
\STATE Estimate $L_u, L_z, L_{\bar{z}}, L_w$ from trajectory pairs
\STATE Compute covering radii $\varepsilon_x, \varepsilon_z, \varepsilon_{\bar{z}}, \varepsilon_w$
\STATE $r \leftarrow L_z \varepsilon_z + L_{\bar{z}} \varepsilon_{\bar{z}} + L_u L_\theta \varepsilon_x + L_w \varepsilon_w$
\STATE $\delta \leftarrow (T_{\max} - T_{\min})/2$
\IF{$r < \delta$ \AND all data points satisfy contracted bounds}
    \RETURN \textbf{CERTIFIED SAFE}, margin $= \delta - r$
\ELSE
    \RETURN \textbf{NOT CERTIFIED}
\ENDIF
\end{algorithmic}
\end{algorithm}

\section{Simulation Results}
\label{sec:results}

This section presents the simulation-based evaluation of the
proposed framework. We analyze energy consumption, thermal comfort,
and safety certification results.

\subsection{Energy Consumption}

\begin{figure}[!t]
\centering
\includegraphics[width=\columnwidth]{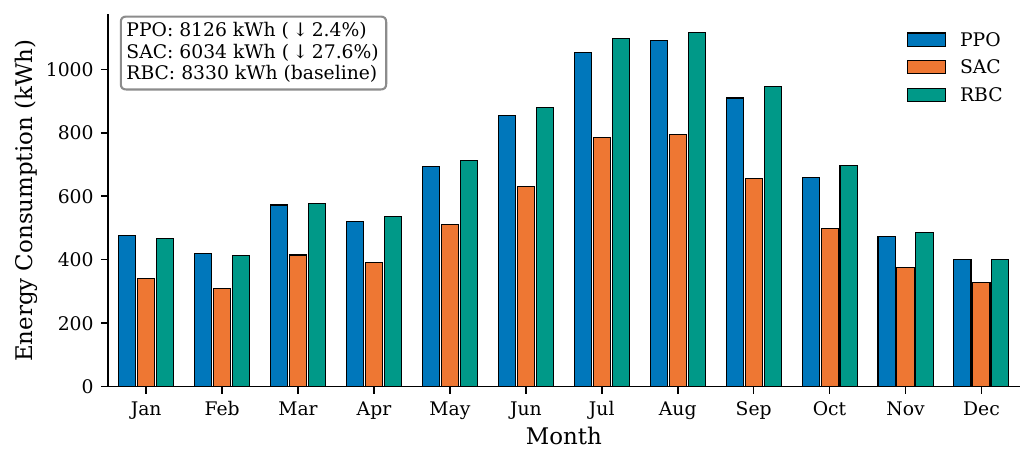}
\caption{Monthly HVAC energy consumption. SAC achieves substantial savings during cooling-dominated months, while PPO maintains consumption similar to RBC.}
\label{fig:energy}
\end{figure}

Fig.~\ref{fig:energy} shows monthly HVAC energy consumption. SAC
achieves the lowest annual consumption at 6034~kWh (27.6\% reduction
vs.\ RBC baseline of 8330~kWh), while PPO consumes 8126~kWh (2.4\%
reduction). SAC's advantage concentrates in cooling-dominated
months (June--September), where it reduces consumption by 35-45\%
through learned pre-cooling strategies. PPO maintains tighter
temperature control, yielding fewer comfort violations but higher
energy use.

\subsection{Thermal Comfort}

\begin{figure}[!t]
\centering
\includegraphics[width=0.8\columnwidth]{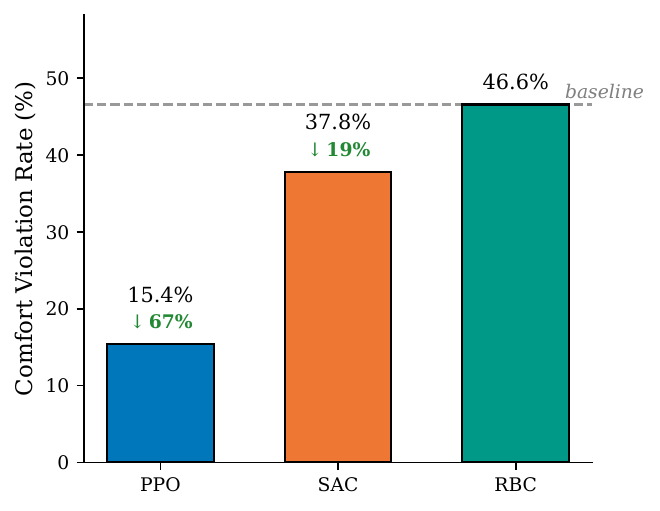}
\caption{Annual comfort violation rates. PPO achieves the lowest violations (15.4\%), outperforming both SAC (37.8\%) and RBC (46.6\%).}
\label{fig:comfort}
\end{figure}

Fig.~\ref{fig:comfort} presents comfort violation rates (percentage of
timesteps with any zone outside $[20, 26]^\circ$C). PPO achieves
15.4\% ($\downarrow$67\% vs.\ RBC at 46.6\%); SAC achieves 37.8\%
($\downarrow$19\%). PPO's clipped surrogate objective constrains
policy updates \cite{schulman2017proximal}, producing conservative policies that prioritize
staying within bounds. SAC's entropy regularization encourages
exploration near comfort boundaries \cite{haarnoja2018soft}, trading violations for energy
savings.

\begin{figure}[!t]
\centering
\includegraphics[width=\columnwidth]{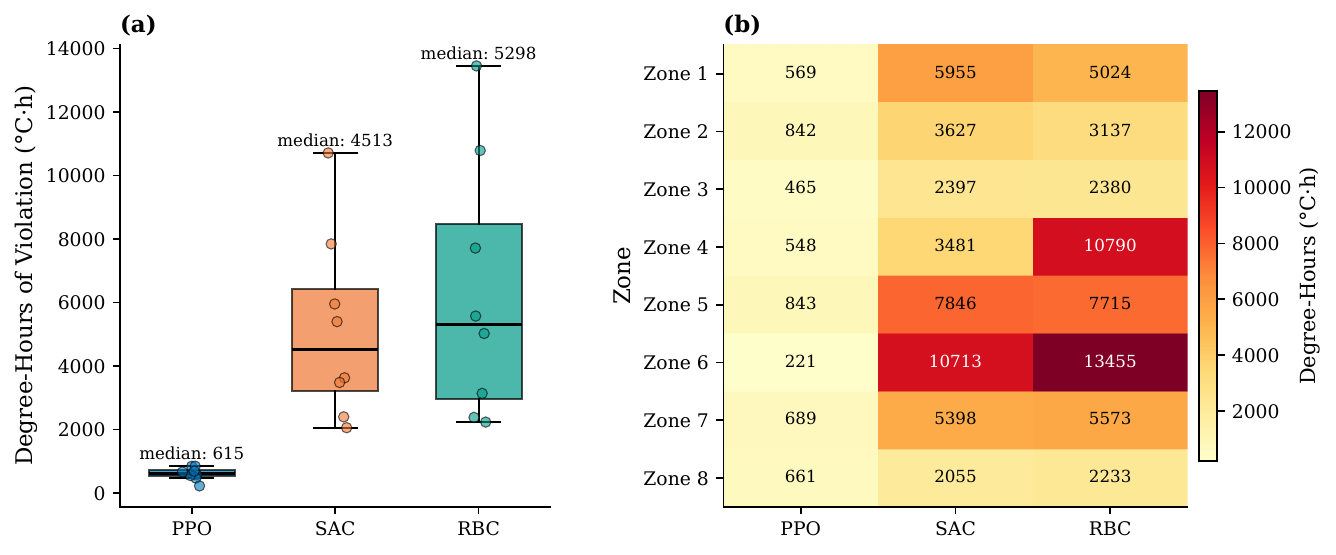}
\caption{Degree-hours of comfort violation. (a) Distribution across controllers. (b) Zone-level breakdown revealing SAC's difficulty with Zone~6.}
\label{fig:degree_hours}
\end{figure}

Fig.~\ref{fig:degree_hours} provides zone-level analysis through
degree-hours of violation. PPO maintains uniform performance across
zones (median: 615$^\circ$C$\cdot$h, IQR: 412--891). SAC shows high
variability (median: 4513$^\circ$C$\cdot$h), with Zone~6 reaching
10713$^\circ$C$\cdot$h due to its high occupant density
(0.50~persons/m$^2$) creating rapid temperature rises that SAC's
energy-minimizing policy fails to anticipate.

\subsection{Energy-Comfort Trade-off}

\begin{figure}[!t]
\centering
\includegraphics[width=0.6\columnwidth]{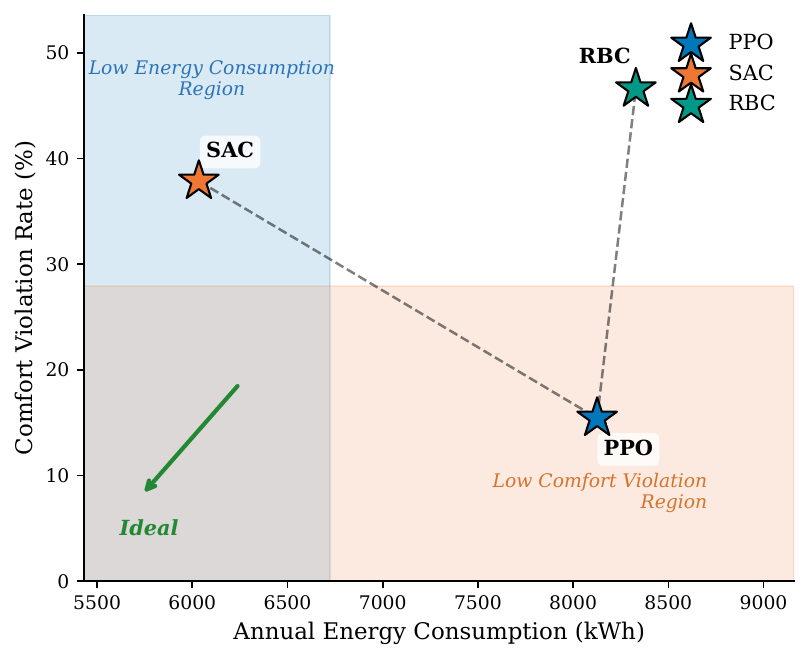}
\caption{Energy-comfort Pareto frontier. PPO and SAC represent distinct operating points; both dominate the RBC baseline.}
\label{fig:pareto}
\end{figure}

Fig.~\ref{fig:pareto} visualizes the Pareto frontier. PPO occupies the
comfort-prioritized region; SAC occupies the energy-prioritized
region. Both dominate the RBC baseline. We note that these
percentage improvements are measured against the RBC; comparison
against a well-tuned MPC or optimized PID baseline would likely yield
smaller margins, which we acknowledge as a limitation.

\FloatBarrier

\subsection{Safety Certification Results}
\label{sec:safety_results}

For safety certification, PPO is retrained with a comfort-biased
reward weight ($\omega = 0.01$, comfort weight $1-\omega = 0.99$).
Table~\ref{tab:lipschitz} reports the Lipschitz constants and covering
radii.

\begin{table}[!t]
\centering
\caption{Lipschitz Constants and Covering Radii}
\label{tab:lipschitz}
\begin{tabular}{lcc}
\toprule
\textbf{Constant} & \textbf{Value} \\
\midrule
$L_\theta$ & 8.298  \\
$L_u$ & 13.517   \\
$L_z$ & 34.616  \\
$L_{\bar{z}}$ & 309.970  \\
$L_w$ & 160.282  \\
\midrule
$\varepsilon_x$ & 0.00296  \\
$\varepsilon_z$ & 0.0107$^\circ$C  \\
$\varepsilon_{\bar{z}}$ & 0.00089  \\
$\varepsilon_w$ & 0.00010  \\
\bottomrule
\end{tabular}
\end{table}

The buffer decomposes as:
$r_z = 0.372^\circ$C (37\%),
$r_{\bar{z}} = 0.277^\circ$C (28\%),
$r_\text{ctrl} = 0.332^\circ$C (33\%),
$r_w = 0.016^\circ$C (2\%),
yielding $r = 0.997^\circ$C. Since
$r < \delta = 3.0^\circ$C, Theorem~\ref{thm:safety} is satisfied with
margin $\delta - r = 2.003^\circ$C. The LipSDP bound is 58\%
tighter than the spectral norm product (19.767).

As $\omega$ increases toward energy prioritization, the policy
explores more aggressively near comfort boundaries, increasing both
the policy Lipschitz constant and the covering radii. Certification
therefore requires a comfort-biased weight; here $\omega = 0.01$
produces a sufficiently smooth policy.

The certification pipeline separates into an offline phase and
an online phase. The offline phase (LipSDP solve, dynamics estimation,
covering radii computation) is executed once after training and does
not affect real-time operation. The online phase reduces to a single
forward pass through the trained policy at each control timestep,
whose computational cost is negligible relative to the 10-minute
sampling interval of the building management system.

\begin{figure}[h!]
\centering
\includegraphics[width=0.51\textwidth]{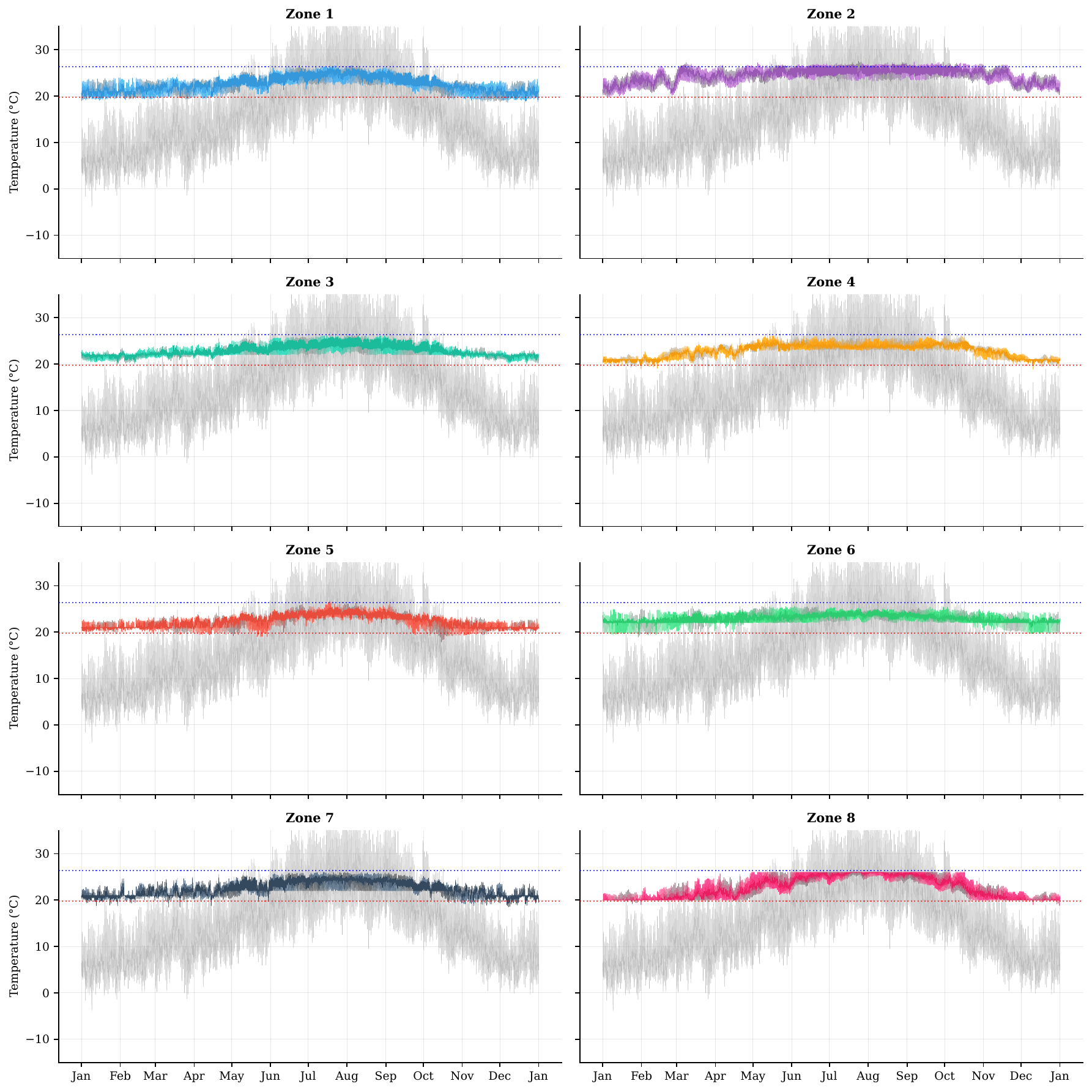}
\vspace{-0.5\baselineskip}
\caption{Annual temperature profiles under PPO control ($\omega = 0.01$). Gray: outdoor temperature. Colored lines: zone air temperatures. Dotted lines: comfort bounds. All zones remain within the certified safe region.}
\label{fig:temp_profiles}
\end{figure}

Fig.~\ref{fig:temp_profiles} validates condition~(ii) of
Theorem~\ref{thm:safety}: all zone temperatures remain within
$[20, 26]^\circ$C throughout the annual cycle.

\section{Conclusion}
\label{sec:conclusion}

This paper presented a safety-certified deep reinforcement learning
framework for multi-zone HVAC control. PPO achieved 67\% reduction in
comfort violations with a certified safety margin of
$2.003^\circ$C, while SAC achieved 27.6\% energy savings. The
channel-wise Lipschitz decomposition separates sensitivity into
temperature, auxiliary, and control channels, yielding tighter
certification bounds than monolithic approaches.

The scope of the current study is shaped by the simulation
environment: EnergyPlus provides a physics-based but idealized VRF
model, and the occupancy profiles follow deterministic schedules
rather than being sampled from a stochastic occupancy model.
The safety certificate relies on Lipschitz constants estimated from
trajectory data, and its validity is tied to the fidelity of these
estimates and the operating conditions they cover.

Future work includes comparison against MPC baselines,
extension to multi-agent architectures with per-zone policy networks,
incorporating safety constraints during training via constrained
policy optimization~\cite{achiam2017constrained}, and validating on physical building systems.

\bibliographystyle{IEEEtran}

\end{document}